\documentclass[11pt,amsmath,amssymb,nofootinbib]{revtex4}
\usepackage{dcolumn}
\usepackage{bm}
\begin{document}
\title{Symplectic geometry and Noether charges for Hopf algebra space-time symmetries}
\author{Michele Arzano}
\email{marzano@perimeterinstitute.ca}
\affiliation{Perimeter Institute for Theoretical Physics\\
31 Caroline St. N, N2L 2Y5, Waterloo ON, Canada}
\author{Antonino Marcian\`o}
\email{antonino.marciano@roma1.infn.it}
\affiliation{Universit\`a degi Studi di Roma ``La Sapienza"\\
Dipartimento di Fisica, P.le A. Moro 2, 00185 Roma, Italy}
\begin{abstract}
\begin{center}
{\bf Abstract}
\end{center}
There has been a certain interest in some recent works in the derivation of Noether charges for Hopf-algebra space-time symmetries. Such analyses relied rather heavily on delicate manipulations of the fields of non-commuting coordinates whose charges were under study. Here we derive the same charges in a ``coordinate-independent" symplectic-geometry type of approach and find results that are consistent with the ones of hep-th/0607221.
\end{abstract}
\maketitle

\section{Introduction}
The emergence of a non-trivial structure of space-time at very short (Planckian) scales has been part of a common intuition in the quantum gravity community since its very early days.  In general the typical new feature expected to emerge at the Planck scale is some sort of ``discretization" in which the Planck length/energy plays a leading role \cite{Garay:1994en}.  Non-commutative geometry is an example of such scenarios where the ``quantization" of classical flat space-time manifold manifests in a non-vanishing commutator for space-time coordinates. In general non-commutative space-times do not enjoy usual relativistic symmetries, Poincar\'e invariance can be either broken or replaced by ``deformed" symmetries described by certain Hopf algebras \cite{Amelino_fate, Majid-Ruegg,Chaichian:2004yh}.\\
In recent years, the possibility that non-classical, ``quantum", symmetries of flat space-time might emerge in the low-energy limit of quantum gravity has also received much attention (see e.g. \cite{Amelino-Camelia:2003xp, Freidel:2005me}). Further interest in these specific type of quantum symmetries, described by the $\kappa$-Poincar\'e Hopf algebra, has been motivated by the suggestion that they might lead to observable effects, e.g. possible experimental signatures of a modification of the energy/momentum dispersion relation in certain astrophysical phenomena (see e.g.\cite{Amelino}).\\  
The relevance of $\kappa$-Poincar\'e in the context of non-commutative geometry relies on the observation \cite{Majid-Ruegg} that such Hopf algebra can be regarded as the symmetry algebra of the so-called $\kappa$-Minkowski non-commutative space-time (NCST). 
One of the major difficulties in working with such a space is that in order to properly handle fields as functions of non-commuting coordinates one has to fix an ordering prescription.  Different choices of ordering prescription give rise to different actions of the symmetries on the algebra of functions which in turn correspond to different choices of the basis of generators of $\kappa$-Poincar\'e.  Such freedom in the choice of ordering leads to {\it apparent} ambiguities in the physical interpretation of such theories \cite{Ale2, Kolno} e.g. the energy and the wave-vector of a field would appear to be related by different dispersion relations depending on the choice of ordering.\\ 
As a contribution to the understanding of these issues, in \cite{knoeth}, in collaboration with Agostini, Amelino-Camelia and Tacchi, we obtained, for the first time, the Noether charges associated to the translation sector of the $\kappa$-Poincar\'e symmetries for a classical field theory on $\kappa$-Minkowski spacetime.  Such charges were shown to be, although for narrow class of non-linear basis redefinitions, invariant by change of basis of the $\kappa$-Poincar\'e Hopf algebra, i.e. to be invariant by ordering convention of the elements of the $\kappa$-Minkowski space-time manifold.  More recently in \cite{Freikono}, working with the same type NCST but considering a different differential calculus and a different formalization of the fields, the authors obtained charges that are different from the ones of \cite{knoeth}. Since these studies relied heavily on different assumptions, we thought it could be useful to develop an alternative procedure, by simplectic-geometry considerations, in analyzing the problem of Noether charges. Indeed a symplectic, ``non-coordinate", formulation \cite{Crnkovic:1986ex} appears to be extremely useful for the purpose of study the conserved charges associated with Hopf algebra symmetries.\\
In our new framework, the Noether charges we derive coincide with the ones obtained in \cite{knoeth}.  Moreover in the formalism introduced it is easy too see that, once a concept of deformed symmetry transformation has been given, these charges are unique.  We hope that such formalism might provide, in general, a powerful tool for the study of classical (and quantum) field theories on non-commutative spaces.\\
In section II A we review the symplectic-geometry approach to classical field theories and adapt it to the Noether analysis for a linear massless scalar field invariant under standard relativistic symmetries.  Such approach, based on the symplectic geometry of the phase space, gives a clearer picture of the symmetry structure of the theory since the charges associated with local symmetries can be expressed in terms of the central object: the symplectic product \cite{Crnkovic:1986ex}.\\
In section II B we summarize the main features of $\kappa$-Poincar\'e ``quantum" relativistic symmetries and the transition from the standard Lie algebra setting to Hopf algebra symmetries.\\
In section III, we show how to define an analogous symplectic structure for a field theory whose symmetries are described by $\kappa$-Poincar\'e.  This will be done introducing a suitable mapping from the space of solutions of the standard equation of motion to the solutions of the $\kappa$-deformed Klein-Gordon equation.  As an application of the new formalism we derive the Noether charges associated with translational symmetries in few, simple steps.\\
Section IV contains some closing remarks.

\section{Preliminaries}
\subsection{Symplectic geometry and phase space of classical field theories}
The idea behind a symplectic geometrical description of classical field theories is to provide a canonical formulation of the phase space which does not spoil the symmetries of the theory, {\it in primis} Poincar\'e invariance.
The central object of such description is a non-degenerate 2-form $\omega$ which provides a ``coordinate-free" global expression for the Poisson brackets.  Such a description of the phase space will turn out very useful for our purposes allowing a clear identification of the Noether charges associated with space-time symmetries when the standard Poincar\'e Lie algebra is replaced by a quantum (Hopf) algebra.\\
We will focus on the simplest example of a massless scalar field with Lagrangian
\begin{equation}
L=\int_{M}\frac{1}{2}(\partial_{\alpha}\Phi\partial^{\alpha}\Phi)
\end{equation}
where $M$ is the space-time manifold on which the theory is defined which in our particular case is just standard Minkowski space-time.  The phase space of such theory is the infinite dimensional manifold $\Gamma$ whose elements are the pairs $\{\Phi,\Pi\}$ where $\Pi=\frac{\partial \mathcal{L}}{\partial \dot{\Phi}}$ is the momentum (density) canonically conjugate to $\Phi$.  Such definition requires singling out a time coordinate $t$, a choice that obscures the relativistic invariance of the theory.  The crucial observation \cite{Crnkovic:1986ex} is that there is indeed a {\it one-to-one} correspondence between the points of the classical phase space and the space of solutions of the equation of motion
\begin{equation}
\label{KleinG}
\Box \Phi=0\, ,
\end{equation}
thanks to the uniqueness properties of the initial value problem for such equation in Minkowski space.  In other words, in our geometrical description, points in phase space are identified with solutions of the classical equation of motion i.e. the phase space $\Gamma$ {\it is} the space of solutions $\mathcal{S}$.
\\
The next step in the geometrical formulation is the definition of a symplectic structure through the introduction of the non-degenerate two-form $\omega$.  To do so one needs to specify what are the functions, tangent and cotangent elements on $\mathcal{S}$.  A function on $\mathcal{S}$ is e.g. the map 
\begin{equation}
\Phi(x):\Phi\in\mathcal{S}\rightarrow\Phi(x)\in\mathbb{R}\, ,
\label{fun1}
\end{equation}
while a tangent vector $\delta \Phi \in T\mathcal{S}$ is a small displacement in $\Phi$ which is compatible with (\ref{KleinG})
\begin{equation}
\label{deltaKleinG}
\Box \delta\Phi=0\, .
\end{equation}
One-forms on $\mathcal{S}$ are linear functionals on  $T\mathcal{S}$.  For every $\delta \Phi \in T\mathcal{S}$ its value at a space-time point $x\in M$ is a real number so we can define a one-form 
$\delta \Phi (x)\in T^{*}\mathcal{S}$ as a mapping 
\begin{equation}
\delta \Phi(x):\delta\Phi\in T \mathcal{S}\rightarrow\delta\Phi(x)\in\mathbb{R}\, .
\end{equation}
A general p-form will be written as \cite{Crnkovic:1986ex}
\begin{equation}
F=\int dx_1\dots dx_p f_{x_1\cdots x_p}\delta \Phi (x_1)\wedge\cdots\wedge\Phi (x_p)
\end{equation}
where the integral is analogous to a sum over the indices $x_1\cdots x_p$ and as usual 0-forms are just functions on $\mathcal{S}$.  One can define an exterior derivative $\delta$ mapping p-forms to p+1-forms, satisfying 
\begin{equation}
\delta^2=0\, ,\,\,\,\,\, \delta(F G)=(\delta F)\, G+(-1)^{r(F)} F\,\delta(G)\, ,
\end{equation}
where $r(F)$ is the rank of $F$, whose action on functions and p-forms is given by
\begin{equation}
\delta f =\int dx \frac{\partial f}{\partial \Phi(x)} \delta \Phi (x)
\end{equation}
\begin{equation}
\delta F=\int dx_1\dots dx_n \frac{\partial f_{x_1\cdots x_n}}{\partial \Phi(x_0)} 
\delta  \Phi(x_0)\wedge\delta \Phi (x_1)\wedge\cdots\wedge\Phi (x_n)\, .
\end{equation}
The non-degenerate, symplectic 2-form $\omega$ will be defined by
\begin{equation}
\label{omega1}
\omega\equiv\delta\left(\int_{\Sigma}d\sigma_{\alpha} J^{\alpha} \right)
\end{equation}
with the ``symplectic potential current" $J^{\alpha}$ defined by
\begin{equation}
J^{\alpha}\equiv\frac{\partial L}{\partial (\Phi _{,\alpha})}\delta \Phi(x)
\end{equation}
where $\Phi _{,\alpha}=\partial_{\alpha}\Phi$ is viewed as an independent field from $\Phi$.  It can be easily shown \cite{Crnkovic:1986ex} that $\omega$ is Poincar\'e invariant and for a vector field tangent to the orbit of space-time symmetry $V_s\in T\mathcal{S}$ 
\begin{equation}
\omega(V_s)=-\delta (Q_s)
\end{equation}
where $Q_s$ is the generator of the symmetry.  In particular for a choice of $\Sigma$ as the standard $t=0$ space-like hypersurface (\ref{omega1}) reduces to the more familiar expression for the symplectic form on the phase space manifold
\begin{equation}
\omega=\frac{1}{2}\int_{\Sigma_t}\delta\Pi \wedge \delta\Phi
\end{equation}
and
\begin{equation}
\omega(V_s)=\frac{1}{2}\int_{\Sigma_t}\left(\delta\Pi(V_s) \delta\Phi-\delta\Pi \delta\Phi(V_s)\right)\, .
\end{equation}
As a vector space, $\mathcal{S}$ is isomorphic to the tangent space at any given point $T_{\Phi}\mathcal{S}$.  Through this correspondence the 2-form $\omega$ and the one form $\omega(V_s)$ can be used to define, respectively, a bilinear and a linear functional on $\mathcal{S}$.  The bilinear will induce a symplectic product on $\mathcal{S}$ 
\begin{equation}
\omega(\Phi_1,\Phi_2)=\frac{1}{2}\int_{\Sigma_t}(\Pi_1\Phi_2-\Phi_1\Pi_2)
\end{equation}
and the linear functional $\omega(V_s)$ evaluated on a solution $\Phi$ will give the (conserved) value of the Noether charge associated with the symmetry
\begin{equation}
\omega(V_s)(\Phi)=\frac{1}{2}\int_{\Sigma_t}\left((\mathcal{L}_{(V_s)}\Pi) \Phi - \Pi (\mathcal{L}_{(V_s)}\Phi) \right)=-Q_s\, 
\end{equation}
where the contraction of a 1-form with the Killing vector field $V_s$ is replaced by the Lie derivative\footnote{From the expression of the Lie derivative of a general p-form $\mathcal{L}_{V_s}F=(\delta F)(V_s)+\delta(F(V_s))$ which for a function (0-form) implies $\mathcal{L}_{V_s}f=(\delta f)(V_s)$.} of $\Phi\in\mathcal{S}$ and its canonically conjugate momentum\footnote{In general, introducing the notion of ``covariant conjugate momenta" \cite{ozaki} $\Pi_{\alpha}=\Phi_{,\alpha}$, the expressions below can be written in covariant form as
\begin{equation*}
\omega(\Phi_1,\Phi_2)=\frac{1}{2}\int_{\Sigma}d\sigma_{\alpha}(\Pi^{\alpha}_1\Phi_2-\Phi_1\Pi^{\alpha}_2)
\end{equation*}
and
\begin{equation*}
\omega(V_s)(\Phi)=\frac{1}{2}\int_{\Sigma}d\sigma_{\alpha}
\left((\mathcal{L}_{(V_s)}\Pi^{\alpha}) \Phi - \Pi^{\alpha} (\mathcal{L}_{(V_s)}\Phi) \right)=-Q_s\, .
\end{equation*}}
 $\Pi$.
When the Killing vector fields correspond to (external) space-time symmetries we will use the following notation for the action of the Lie (Hopf) algebra element associated with the symmetry on functions
\begin{equation}
\mathcal{L}_{(M_{\mu\nu})}\Phi=M_{\mu\nu}\vartriangleright \Phi\, 
\end{equation}
to stress the fact that $\mathcal{S}$ is an $\infty$-dimensional representation of the Poincar\'e algebra.\\
On the space of complex solutions of the (massless) Klein-Gordon equation $\mathcal{S}^{\mathbb{C}}$ our symplectic structure will define an hermitian inner product 
\begin{equation}
(\Phi_1,\Phi_2)=-2i\,\omega(\Phi^*_1,\,\Phi_2)
\end{equation}
where $\Phi^*_i$ denotes the complex conjugate of $\Phi\in\mathcal{S}^{\mathbb{C}}$.  It will be useful for our purposes to write such product in an explicit covariant form as 
\begin{equation}
(\Phi_1,\Phi_2)=\int\frac{d^4 p}{(2\pi)^3}\delta(\mathcal{C}(p))\frac{p_0}{|p_0|}\tilde{\Phi^*}_1{(-p)}\tilde{\Phi}_2(p)
\end{equation}
where $\tilde{\Phi}_i(p)$, $\tilde{\Phi^*}_i(p)$ are the Fourier transforms of $\Phi_i$, $\Phi^*_i$ and $C(p)$
is the relativistic ``mass" Casimir which relates the Fourier parameters $p_{\mu}$ according to (\ref{KleinG}).\\
Notice how all the formulas above are independent of the choice of basis of functions on $\mathcal{S}$.  
In this letter will focus on the Noether charges associated with the space-time translation symmetries. As we stressed in the previous section the latter can be expressed in terms of the action of the symmetry algebra generators $P_{\mu}$ acting on the fields $\Phi$ and ``freezing" such action, in a coordinate independent way, using the conserved symplectic product.  We have
\begin{equation}
Q_{\mu}=\frac{1}{2}(\Phi, P_{\mu} \vartriangleright \Phi)\, 
\end{equation}
which written in a explicit form reads 
\begin{equation}
Q_{\mu}= \int\frac{d^4 p\,}{2(2\pi)^3}\,\,\delta(\mathcal{C}(p))\,\,\frac{p_0}{|p_0|}
\,\,p_{\mu}\,\,\tilde{\Phi^*}(-p)\tilde{\Phi}(p)\, . 
\end{equation}

\subsection{Relativistic symmetries: from (Poincar\'e) Lie algebras to Hopf algebras}
As mentioned in the introduction, the idea that Hopf algebras might replace the standard framework of Lie algebras in describing the symmetries of flat space-time, as it emerges from the low-energy limit of a general theory of quantum gravity, has become increasingly popular in recent years. \\ 
It is well known that the standard Poincar\`e algebra $\mathcal{P}$  can be endowed with the structure of a (trivial) Hopf algebra.  If $Y=\{P_{\mu},M_{\mu\nu}\}$ are the generators of $\mathcal{P}$, the co-product, co-unit and antipode can be defined respectively as
\begin{equation}
\Delta(Y)=Y\otimes 1+1\otimes Y,\,\,\,\,
\epsilon(Y)=0,\,\,\,\,S(Y)=-Y\, .
\end{equation}
We will consider $\kappa$-deformations of relativistic symmetries described by a Hopf algebra in which both the co-algebra sector and the product (commutator) of the algebra elements (symmetries generators) are deformed.  In particular, in this letter, we will focus on the example of the $\kappa$-Poincar\`e Hopf algebra $\mathcal{P}_{\kappa}$ \cite{Lukierski:1992dt}.  In the {\it bicrossproduct basis} \cite{Majid-Ruegg} the deformed commutators for the generators of translations $P_0,P_i$, rotations $M_i$ and boosts $N_i$ are
\begin{eqnarray}
&[P_{0},P_{j}]=0 \qquad [M_j,M_k]=i \epsilon_{jkl}M_l \qquad [M_j,N_k]=i \epsilon_{jkl}N_l  \qquad [N_j,N_k]=i \epsilon_{jkl}M_l  \nonumber\\
&[P_0,N_l]=-iP_l \qquad [P_l,N_j]=-i\delta_{lj}\Big( \frac{\kappa}{2}  \left(1-e^{-\frac{2 P_0}{\kappa}} \right) +\frac{1}{2 \kappa} \vec{P}^2 \Big)+  \frac{i}{\kappa}P_l P_j \nonumber \\
&[P_0,M_k]=0 \qquad [P_j,M_k]=i \epsilon_{jkl}P_l
\end{eqnarray}\\
where $\epsilon_{jkl}$ is the Levi-Civita symbol and  $j,k,l$ are spatial indices.\\
The non-cocommutative coproducts are
\begin{eqnarray}
\Delta(P_0)&=&P_0\otimes 1+1\otimes P_0\,\,\,\,\,\,\Delta(P_j)=P_j\otimes 1+e^{-P_0/\kappa}\otimes P_j \nonumber \\
\Delta(M_{j})&=&M_{j}\otimes 1+1\otimes M_{j} \nonumber \\
\Delta(N_j)&=&N_j\otimes 1+e^{-P_0/\kappa}\otimes N_j+\frac{\epsilon_{jkl}}{\kappa}P_k\otimes N_l\, . \label{copro}
\end{eqnarray}\\
the antipodes 
\begin{eqnarray}
S(M_l)&=&-M_l \nonumber\\
S(P_0)&=&-P_0 \nonumber\\
S(P_l)&=&-e^{\frac{P_0}{\kappa}}P_l \nonumber\\
S(N_l)&=&-e^{\frac{P_0}{\kappa}}N_l+\frac{1}{\kappa}\epsilon_{ljk}e^{\frac{P_0}{\kappa}}P_j M_k  \label{anti}\, ,
\end{eqnarray}
while for the co-units 
\begin{equation}
\epsilon(P_{\mu})=\epsilon(M_j)=\epsilon(N_k)=0.
\end{equation}
The (deformed) mass Casimir $C_{\kappa}$ of the $\kappa$-Poincar\'e Hopf algebra links the  generators of translations and is given by 
\begin{equation}
C_{\kappa}=\left( 2\kappa \sinh \left( \frac{P_0}{2 \kappa}\right)  \right)^2-\vec{P}^2e^{\frac{P_0}{\kappa}}.  
\end{equation}

\section{Noether charges for $\kappa$-Poincar\'e Hopf algebra symmetries}
In this section we consider a classical field theory with space-time symmetries described by the non-trivial Hopf algebra introduced above.  We will see how, in analogy with the standard case, the phase space of the theory can be described in terms of the space of solutions of the deformed equation of motion.  It will be shown how such a space, as a vector space, is isomorphic to the space of solutions of the standard equations of motion.  Such isomorphisms will enable us to introduce a conserved inner product on the $\kappa$-space of solutions and we shall use such product to express the (conserved) Noether charges associated with space-time translation operators providing a simplified derivation of the results obtained in \cite{knoeth}.\\
In our Hopf algebra symmetry case plane waves
will still be solutions of a deformed Klein-Gordon equation \cite{kDirac}.  However wave exponentials labeled by a given value of the spatial momentum will combine according to the coproduct structure of the non-trivial co-algebra sector of the Hopf algebra. Seen from a dual space-time point of view \cite{ArzAme} this corresponds to introducing a $*$-multiplication for the algebra of functions on space-time coordinates.  
Roughly speaking any such $*$-structure corresponds to a different choice of normal ordering for functions of non-commuting coordinates or a different choice of basis for the generators of the $\kappa$-Poincar\'e (Hopf) algebra \cite{Ale2}.  Unlike the standard Lie-algebra symmetry case, in $\kappa$-Poincar\'e in the bicrossproduct basis, for each mode $\vec{p}$ the positive and negative \underline{real} roots of the deformed mass Casimir $\mathcal{C}_{\kappa}(p)$ will not be equal in modulus.  Infact there will be two real roots\footnote{In $\mathbb{C}$ we will actually have a countable infinite set of (complex) roots for $\mathcal{C}_{\kappa}(p)=0$. As it will become clear from our considerations below, each pair of such complex roots will correspond to a copy of the same deformed space of solutions.  For our purposes the restriction to a single pair of (real) roots will be enough to construct a model of field theory with {\it deformed} relativistic symmetries.}:
\begin{equation}
\label{kroots}
p^0_{\pm}(\vec{p})=\kappa\log(\Omega(\vec{p})_{\pm})\,
\end{equation}
where $\Omega_{\pm}(\vec{p})$ is a  function of $|\vec{p}|$ and in our specific case reads
\begin{equation}
\Omega(\vec{p})_{\pm}=\frac{1}{1\mp\frac{|\vec{p}|}{\kappa}}
\end{equation}
In the limit $\kappa\rightarrow\infty$ the real part of each solution approaches the value of the standard roots $p^0_{\pm}=\pm|\vec{p}|$.\\
We denote with  $\mathcal{S}^{\mathbb{C}}_{\kappa}$ the spaces of (complex) solutions of the $\kappa$-Klein-Gordon equation.  
The set of plane waves $\left\{\phi_{\vec{p} \pm}^{\kappa}\right\}$, with the subscript $\vec{p} \pm$ specifying plane waves whose momentum labels are on the positive or negative deformed mass-shell, will provide a basis for the associated vector space $\mathcal{S}^{\mathbb{C}}_{\kappa}$.  In fact such functions are linearly independent (i.e. every finite sub-collection is linear independent) and they span $\mathcal{S}^{\mathbb{C}}_{\kappa}$.  This can be easily seen introducing a linear map 
\begin{equation}
\mathfrak{m}:\mathcal{S}^{\mathbb{C}}_{\kappa} \rightarrow \mathcal{S}^{\mathbb{C}} 
\end{equation}
which associates every element of $\mathcal{S}^{\mathbb{C}}_{\kappa}$ to its ``classical" counterpart in $\mathcal{S}^{\mathbb{C}}$.
In particular we have for a basis of plane wave solutions
\begin{equation}
\mathfrak{m}(\phi_{\vec{p}\pm}^{\kappa}) = \phi_{\vec{p}\pm}\, ,
\end{equation}
where, with obvious notation, $\phi_{\vec{p}\pm}$ are standard plane wave solutions labeled  by points on the positive or negative undeformed mass-shell $p^0_{\pm}=\pm|\vec{p}|$.
The map $\mathfrak{m}$ is an isomorphism of vector spaces (linear, one-to-one and onto), which proves the above claim.
Notice here that, in analogy with the undeformed case, there will be two subspaces of $\mathcal{S}^{\mathbb{C}}_{\kappa}$: $\mathcal{S}^{\mathbb{C}}_{\kappa+}$ and $\mathcal{S}^{\mathbb{C}}_{\kappa-}$, spanned respectively by $\left\{\phi_{\vec{p}+}^{\kappa}\right\}$ and $\left\{\phi_{\vec{p}-}^{\kappa}\right\}$.  These spaces are mapped into each other by {\it deformed} complex conjugation 
\begin{equation}
\bar{\phi}_{\vec{p}_+}^{\kappa}=\phi_{\dot{-}\vec{p}_+}^{\kappa}=\phi_{\vec{k}_-}^{\kappa}
\end{equation}
with $\vec{k}$ given by $\vec{k}=-\vec{p}e^{\frac{p_0^+}{\kappa}}$ and the label $\dot{-}\vec{p}_{\pm}$ is a short hand notation meaning that we are taking the antipode (\ref{anti}) of a four-momentum on-shell in the plane wave.\\
Both in the standard and in the $\kappa$-deformed case the spaces of positive and negative energy solutions can be shown to be isomorphic to the spaces of functions on the positive and negative mass-shell, respectively, through (non-commutative) Fourier transforms \cite{Ale2}.  In analogy with a Weyl map \cite{Ale} for functions of spacetime coordinates, using the map $\mathfrak{m}$ one can associate every element in the (commutative) algebra of functions of standard solutions with elements of a non-commutative algebra of functions of deformed solutions.  Indeed, using the notation introduced in (\ref{fun1}) for functions on $\mathcal{S}$, we can write
\begin{equation}
(\Phi(x)\circ\mathfrak{m})\cdot(\Phi(y)\circ\mathfrak{m})=(\Phi(x)*\Phi(y))\circ\mathfrak{m}
\end{equation}
where $\Phi(x)$, $\Phi(y)$ are two distinct functions on $\mathcal{S}$ and $*$ is the non-commutative multiplication which turns the space of functions on $\mathcal{S}$ into a non-commutative algebra.\\
As a vector space $\mathcal{S}^{\mathbb{C}}_{\kappa}$ can be endowed with a symplectic structure if we require  $\mathfrak{m}$ to be a {\it Poisson map}.  More precisely we have that in general a map between Poisson manifolds $M:P\rightarrow Q$, it is said to be a Poisson map iff
\begin{equation}
\{f_1,f_2\}_Q\circ M=\{f_1\circ M,f_2\circ M\}_P
\end{equation}
with $f_1,f_2$ smooth functions on $Q$.  Now suppose we go back for a moment to the space of ``standard" real solutions $\mathcal{S}$.  We have a map $\mathfrak{m}:
\mathcal{S}_{\kappa}\rightarrow\mathcal{S}$. Functions on $\mathcal{S}$ can be defined in analogy to Section II A and, keeping in mind the isomorphism between $\mathcal{S}$ and the space of functions on the mass-shells, as
\begin{equation}
\Phi(\vec{p}):\Phi\in\mathcal{S}\rightarrow\tilde{\Phi}(\vec{p})\in\mathbb{R}
\end{equation}
where $\tilde{\Phi}(\vec{p})$ has support on the positive or negative undeformed mass-shells. 
Tangent vectors $\delta\Phi\in T\mathcal{S}$ are small displacements in the field configuration (see Section II A).  We have
\begin{equation}
\{\Phi(\vec{p}_1),\Phi(\vec{p}_2)\}=\omega(\delta\Phi_1,\delta\Phi_2)
\end{equation}
from which, as we stressed in Section II A, one can define a bilinear functional $\omega(\Phi_1,\Phi_2)$ on $\mathcal{S}$.  Our Poisson map induces a symplectic structure on $\mathcal{S}_{\kappa}$ through
\begin{equation}
\{\Phi(\vec{p}_1)\circ\mathfrak{m},\Phi(\vec{p}_2)\circ\mathfrak{m}\}_{\kappa}=
\{\Phi(\vec{p}_1),\Phi(\vec{p}_2)\}\circ\mathfrak{m}\,.
\end{equation}
This, in turn, corresponds\footnote{Fixing a point $\Phi\in\mathcal{S}_{\kappa}$ the condition for $\mathfrak{m}$ to be a Poisson map translates into the following condition on the symplectic 2-forms $\omega$ and $\omega_{\kappa}$ 
\begin{equation*}
(\mathfrak{m}_{\Phi})_{*}\otimes(\mathfrak{m}_{\Phi})_{*}(\omega)_{\mathfrak{m}(\Phi)}=(\omega_{\kappa})_{\Phi}
\end{equation*}
with $(\omega_{\kappa})_{\Phi}\in\Lambda^2_{\Phi}\mathcal{S_{\kappa}}$ and $(\omega)_{\mathfrak{m}(\Phi)}\in\Lambda^2_{\mathfrak{m}(\Phi)}\mathcal{S}$ and where $(\mathfrak{m}_{\Phi})_{*}:T^*_{\mathfrak{m}(\Phi)}\mathcal{S}\rightarrow T^*_{\Phi}\mathcal{S}_{\kappa}$ is the pullback associated with $\mathfrak{m}$.} 
to a bilinear functional on $\mathcal{S}_{\kappa}$ given by
\begin{equation}
\omega_{\kappa}(\Phi^{\kappa}_1,\Phi^{\kappa}_2)=(\mathfrak{m}^{-1}\otimes\mathfrak{m}^{-1})
(\omega(\Phi_1\,,\Phi_2)).
\end{equation}
Going back to the complexified spaces $\mathcal{S}^{\mathbb{C}}_{\kappa}$ and $\mathcal{S}^{\mathbb{C}}$ we can write down an explicit form of the deformed hermitian product\footnote{Here we drop the superscript $\kappa$ being now obvious that in the following we refer to elements of $\mathcal{S}^{\mathbb{C}}_{\kappa}$.} 
$(\Phi_1,\Phi_2)_{\kappa}=-2i\omega_{\kappa}(\Phi^*_1,\Phi_2)$ as
\begin{equation}
\label{kinnerp}
(\Phi_1,\Phi_2)_{\kappa}=\int\frac{d^4 p}{(2\pi)^3}\,\,\delta(\mathcal{C}_{\kappa}(p))\,\,\frac{p_0}{|p_0|}
\,\,e^{\frac{3p_0}{\kappa}}\,\tilde{\Phi^*}_1(\dot{-}p)\tilde{\Phi}_2(p)\, ,
\end{equation}
where $C_{\kappa}(p)$, the deformed mass Casimir of $\kappa$-Poincar\'e, describes the deformed mass-shell and we used the fact the following property $\tilde{\Phi^*}(\dot{-}p)=e^{-\frac{3p_0}{\kappa}}\left(\tilde{\Phi}(p)\right)^*$  holds for $\kappa$-Fourier transforms \cite{Ale2}.\\
We can now associate conserved quantities to the symmetries of the theory, in particular we can derive the energy and momentum charges associated with the $\kappa$-deformed translation symmetries for each $\Phi\in\mathcal{S}^{\mathbb{C}}_{\kappa}$
\begin{equation}
Q^{\kappa}_{\mu}=\frac{1}{2}(\Phi, P^{\kappa}_{\mu} \vartriangleright \Phi)_{\kappa}\,
\end{equation}
which written in explicit form is analogous to the result found in \cite{knoeth}
\begin{equation}
Q^{\kappa}_{\mu}= \int\frac{d^4p}{2(2\pi)^3}\,\,\delta(\mathcal{C}_{\kappa}(p))\,\,\frac{p_0}{|p_0|}
\,\,p_{\mu}\,\,e^{\frac{3p_0}{\kappa}}\,\tilde{\Phi^*}(\dot{-}p)\tilde{\Phi}(p)\, . \label{Heine}
\end{equation}
Each choice of map $\mathfrak{m}$ amounts to a choice of formalization of the same classical field theory and to the same class of deformed transformations.  The point we would like to stress here is that the approach we followed naturally leads to a unique form of deformed Noether charges associated with these new (Hopf algebra) translation symmetries.

\section{Conclusions}
We have introduced a new (symplectic) geometrical approach to study classical field theories on non-commutative spaces invariant under deformed (Hopf algebra) symmetries.  The new formalism provides a clear picture of how invariant quantities can be associated with the symmetries of the theory.  In particular we were able to derive the Noether charges relative to deformed translations in a rather simple way and see how they are univocally determined once a notion of deformed symmetry transformation has been specified. 

\begin{acknowledgments}
We are very grateful to Giovanni Amelino-Camelia for stimulating discussions during the preparation of the present work.  We would also like to thank Bianca Dittrich, Laurent Freidel and Jurek Kowalski-Glikman for helpful remarks.   AM would like to thank Perimeter Institute for hospitality while part of this work was being completed.\\
Research at Perimeter Institute for Theoretical Physics is supported in
part by the Government of Canada through NSERC and by the Province of
Ontario through MRI.
\end{acknowledgments}

\end{document}